# Automated Fourier space region-recognition filtering for off-axis digital holographic microscopy

Xuefei He[1], Chuong Vinh Nguyen [1,2], Mrinalini Pratap[3], Yujie Zheng[1], Yi Wang[1], David R. Nisbet[1], Richard J Williams[5],Melanie Rug[4], Alexander G. Maier[3], Woei Ming Lee[1] [†]

[1]*Research School of Engineering, College of Engineering and Computer Science, The Australian National University, Canberra ACT 2601, Australia*
[2] *ARC Centre of Excellence for Robotics Vision, College of Engineering and Computer Science, The Australian National University, Canberra ACT 2601, Australia*
[3] *Research School of Biology, College of Medicine, Biology and Environment, The Australian National University, Canberra ACT 2601, Australia*
[4]*Centre for Advanced Microscopy, ANU College of Physical & Mathematical Sciences, The Australian National University, Canberra, ACT 2601, Australia*
[5] *School of Aerospace, Mechanical and Manufacturing Engineering and the Health Innovations Research Institute, RMIT University, Melbourne, Australia*
[†]steve.lee@anu.edu.au

**Abstract:** Automated label-free quantitative imaging of biological samples can greatly benefit high throughput diseases diagnosis. Digital holographic microscopy (DHM) is a powerful quantitative label-free imaging tool that retrieves structural details of cellular samples non-invasively. In off-axis DHM, a proper spatial filtering window in Fourier space is crucial to the quality of reconstructed phase image. Here we describe a region-recognition approach that combines shape recognition with an iterative thresholding to extracts the optimal shape of frequency components. The region recognition technique offers fully automated adaptive filtering that can operate with a variety of samples and imaging conditions. When imaging through optically scattering biological hydrogel matrix, the technique surpasses previous histogram thresholding techniques without requiring any manual intervention. Finally, we automate the extraction of the statistical difference of optical height between malaria parasite infected and uninfected red blood cells. The method described here pave way to greater autonomy in automated DHM imaging for imaging live cell in thick cell cultures.

______________________________________________________________________________

## 1. Introduction

Phase contrast microscopy [1] enables the transformation of phase into intensity that allows clear observation of cellular boundaries, but remains a qualitative imaging tool. The emergence of quantitative phase imaging aims to extract information of the optical height of biological samples. Off-axis holography [2] operates by projecting two equal copies of light into two separate paths, one undisturbed and the other perturbed (phase delay) by a sample or an object. The two paths combine to reveal the amount of phase delay that is recorded on a photosensitive film. Phase delay represents the optical path length difference between the two copies of the beams, which is then used to retrieve the optical height of the object. Combining holography with microscopy and digital imaging devices (charge coupled or complementary metal-oxide), Digital holographic microscope [3, 4] records interference pattern (hologram) that emerges from superposition of a reference wavefront with that emerging from the microscopic sample. Both optical height and refractive index fluctuations are then used to quantify differences between biological samples [5-9]. DHM has been utilized in many biological applications such as determination of cellular motility[5, 10], morphology[11] and biomechanics[12-14].

DHM heavily relies on numerical calculations to accurately reconstruct the phase of the biological sample. The numerical steps used in off-axis DHM technique include: discrete 2D Fourier transform, spatial frequency filtering [15], numerical propagation, phase unwrapping and aberration correction [16]. To achieve high throughput diseases diagnosis, these numerical calculations should be programmed to be automated and adaptive to different sample and imaging conditions. The complexity lies in the spatial frequency filtering process. Since the distribution of spatial frequencies of different samples varies significantly, the filter window needs to change correspondingly for the precise extraction of the desired frequency orders (first orders). One straightforward filtering method for precise extraction is manually selecting the first order [16] in spatial frequency domain. Obviously, this approach requires manual input for each hologram, which is therefore not suitable for automation. While there are some histogram analysis techniques [17, 18] aiming to provide adaptive fitering that could possibly provide automated phase retrieval. Unfortunately, the histogram analysis process still require manual intervention (e.g. initial windowing parameter) to calculate the appropriate threshold. Hence, these techniques is still unable to provide sufficient flexibility or simplicity for automated selection of the appropriate spatial filter for a wide variety of samples especially confluent live cell cultures with highly scattering backgrounds.

Here we propose an adaptive filtering process based on iterative thresholding and region-based selection. This combination gradually selects the optimal frequency component boundary and using shape recognition to extact the optimal frequency component for different holograms. We find that this approach firstly avoids fixing a single thresholding level and removes the need for any manual intervention and any initial input parameter. In addition, the automated adaptive thresholding approach operates successfully with a range of samples including scattering samples when compared to previous techniques[17, 18] (histogram-based). The region recognition approach also demonstrates to be robust for all standard experimental conditions i.e. field of view, sample density, orientation and spacing of interference fringes. The automated DHM system was showed to extract quantitative phase measurements of biological red blood cells (RBCs) infected and uninfected with malaria parasite. All the processes are conducted in a single loop, which allows the continuous imaging and controlled use of the user interface. In the following sections, we explain the steps taken to implement the region recognition and iterative thresholding process as well as the automated sample classification of RBCs.

## 2. Method

Interferometry measures the phase of cross-correlations between the sample and reference wave, shown as following

$$O(\vec{r},t) = |O| exp\left[ j\left(-\omega t + \left(\vec{k}\cdot\vec{r}\right) + \varphi\right)\right]. \quad (1)$$

$$R(\vec{r},t) = |R_0| exp\left[ j\left(-\omega t + \left(\vec{k}\cdot\vec{r}\right)\right)\right]. \quad (2)$$

where $O(\vec{r},t)$ and $R(\vec{r},t)$ are wave functions of object wave and reference wave, $\omega$ indicates the angular frequency of both waves and $\vec{k}\cdot\vec{r}$ indicates the dot product of wave vector $\vec{k}$ ( two waves interfere with each other with same incident angle) and optical length $\vec{r}$. $\varphi$ is the phase delay produced by the sample. The phase change derived from the sample ($O$) that is recorded as a shift in the intensity pattern of fringes I,

$$I = |O + R|^2 = |R|^2 + |O|^2 + OR^* + RO^*. \quad (3)$$

By substituting Eq. (1) and Eq. (2) into Eq. (3), we have

$$I = O_0^2 + R_0^2 + |OR|exp(j\varphi) + |RO|exp(-j\varphi). \tag{4}$$

Based on Eq. (4), the phase shift is recorded in two spatial frequency components $exp(j\varphi), exp(-j\varphi)$ that are spatially overlapping in an inline holography. In off-axis system, the sample and reference beams interfere at a slightly different angle θ creating two separate wave vector $\vec{k}_r, \vec{k}_o$ in two waves:

$$R(\vec{r},t) = |R_0|exp\left[j\left(-\omega t + (\vec{k}_r \cdot \vec{r})\right)\right], \tag{5}$$

$$O(\vec{r},t) = |O|exp\left[j\left(-\omega t + (\vec{k}_o \cdot \vec{r}) + \varphi\right)\right]. \tag{6}$$

Hence the interference pattern is rewritten as:

$$I = O_0^2 + R_0^2 + |OR|exp\left(j(\varphi + (\vec{k}_o - \vec{k}_r) \cdot \vec{r})\right) + |RO|exp\left(-j(\varphi + (\vec{k}_r - \vec{k}_o) \cdot \vec{r})\right) \tag{7}$$

The phase shifts are spatially separated $(\vec{k}_o - \vec{k}_r) \cdot \vec{r}$ and $-(\vec{k}_o - \vec{k}_r) \cdot \vec{r}$ in the frequency domain that makes it possible for single shot phase recovery and high speed imaging [19, 20].

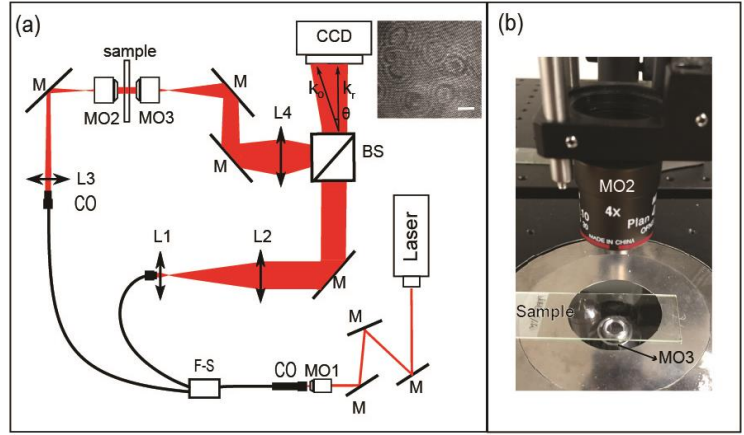

FIG. 1. DHM Imaging setup. (a) L1 and L2 expand the reference beam. L3 focuses object beam onto the back focal plane of a second microscope objective MO3. L4 is a tube lens. M are sliver-coated reflective mirrors to fold the beam along the imaging system. FS is fiber splitter and BS is the beam splitter for splitting and recombining the sample and reference beam. MO1 is the fiber coupling objective, MO2 is the illuminating objective and MO3 is the imaging objective (inset shows is an example of a recorded hologram (scale bar = 5 µm). (b) Image showing the sample placed between the two objective lenses in the setup.

Figure 1(a) shows an off-axis DHM setup that we built on an inverted microscope system (OLYMPUS IX70) using a continuous wave visible laser (λ=632.8nm, Spectra Physics Model 127-3502 Stabilite Polarized Helium-Neon Laser). The output beam is first coupled into a single-mode optical fiber using a microscopy objective (MO1, numerical aperture (N.A. = 0.25), then divided into two separate beams through an optical fiber splitter (F-S, 1 x 4 single mode fiber optic couplers P/N: FCQ632-APC). The outputs of the fibre are collimated using fibre collimators (CO). The beam travelling through the sample [object beam $O(\vec{r},t)$, Eq. (6)] is focused by L3 (f = 200mm) onto the back focal plane of a second microscope objective (MO2, N.A. = 0.1) and collected by an imaging objective lens (MO3). There are two imaging objective lenses (MO3) used in our setup with N.A. = 1.25 and low N.A. = 0.25. The reference beam [sample beam R in Eq. (5)] is collimated and then expanded by lenses L1 (f =

30mm) and L2 (f = 100mm) and finally recombined with the sample beam by a non-polarizing beam splitter (BS, THORLABS CM-BS013) onto a charge coupled device (CCD) camera (Retiga-2000R Fast 1394 Color Cooled) by tube lens L8 built in microscope frame. Figure 1(b) shows experimental setup where the sample is placed in an inverted microscope system. The incident directions of reference and sample beams at the CCD plane are indicated by $\vec{k}_r$ and $\vec{k}_o$ in Fig. 1(a). $\vec{k}_o$ is added at a slight tilted angle θ which results in off-axis holograms digitally recorded as shown in Fig. 1(a) beside the CCD.

## 3. Region-recognition of spatial frequency

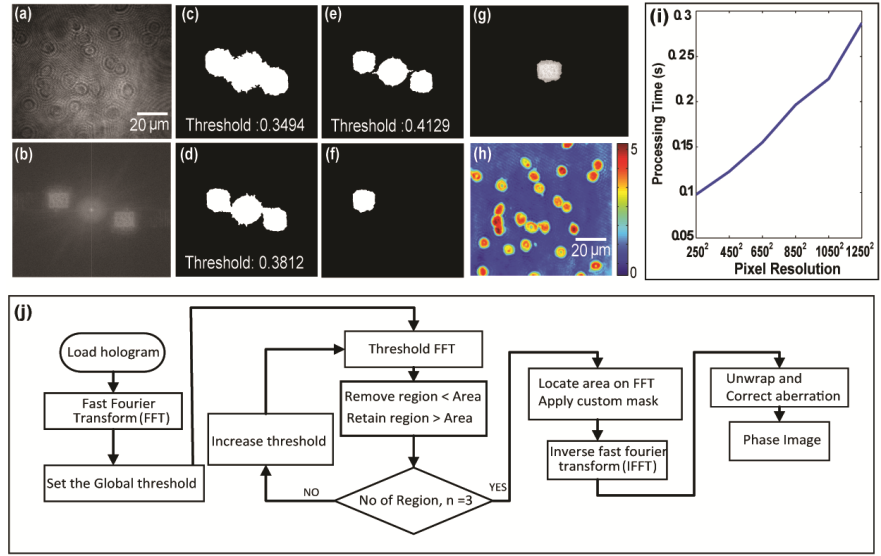

FIG. 2. The region-recognition method. (a) Recorded hologram of RBCs (b) Spatial frequency of the hologram after fast Fourier transform (FFT). (c), (d) and (e), shows intermediate iterative thresholding process with incremental threshold level. (f) Optimised spatial filter window (g) Final filtered and centred spatial frequency area. (h) Reconstructed phase image. The color bar gives the phase values by unit of radian. (i) Calculated time with different pixel size by this region-recognition method. (j) Overview of the flow chart of iterative thresholding and region-recognition program.

After fast Fourier transform (FFT) of the recorded hologram [Fig. 2(a)], the phase shift is recorded in the spatial frequency on two symmetrical areas (+1 and -1 terms) in the frequency domain (FFT hologram). The shapes and sizes of these two symmetrical areas can vary according to different imaging conditions. Figure 4 shows some examples that different samples and imaging conditions give different symmetrical frequency components outlines. This implies that the frequency component needs to be carefully selected for accurate reconstruction. Here we propose an adaptive filtering process that is robust to all standard experimental conditions. The method is built upon an iterative thresholding and region-based selection. The procedures is broken down into 3 steps:

(1) Apply global threshold level (GTL)[21] to the intensity of FFT hologram to get the binary image and then implement the region recognition process of it (*regionprops* MATLAB is used to provide three properties i.e. size of objects, centroid of objects in binary image and the box boundaries; *graythresh* MATLAB is used to get the GTL).
(2) Increase the threshold level by one percent of GTL and repeat the first step until the number of regions reaches three (one for background intensity, one for the desired frequency component and one for the mirrored frequency component).

(3) Use box boundary data from *regionprops* function and the binary image form first step to get the right frequency component boundary and use it as a filtering window (additional Gaussian function is applied to smooth the edge of final filtering window).

Figure 2 shows an example of the adaptive filtering process. Figure 2(a) shows the recorded hologram of RBCs and Fig. 2(b) is the spatial frequency after FFT. Figures 2(c), (d) and (e) show the iterative threshold process with incremental threshold level. The processing time of this method is shown in Fig. 2(i). Once the appropriate spatial frequency is selected and centered [Fig. 2(g)], an inverse FFT (iFFT) is performed to retrieve the complex amplitude of the sample, which can be seen from Fig. 2(h). Figure 2(j) shows the flow chart of iterative thresholding and region-recognition program. Since the phase map acquired from the recorded hologram is limited from $-\pi$ to $\pi$, so called wrapped, the true continuous phase value corresponding to the optical height of cells needs to be unwrapped [22]. Then we apply Zernike polynomials to compensate the aberration [23] in the system.

There have been previous techniques [17, 18] aimed at providing adaptive filtering methods by using histogram analysis to settle the appropriate threshold. However, these methods entail a series of steps that require manual settings or prior knowledge of the frequency distribution. For example, the first procedure in pervious technique [18] is to apply a negative Laplacian [24] to suppress zero order in FFT hologram, which requires distance between first and zero order frequency components to be known prior or well-defined in the FFT hologram. Intuitively, the approach may not operate well when positions of the spatial frequencies change (i.e. different samples and interference fringes on the CCD camera or that the samples has large optical scattering).

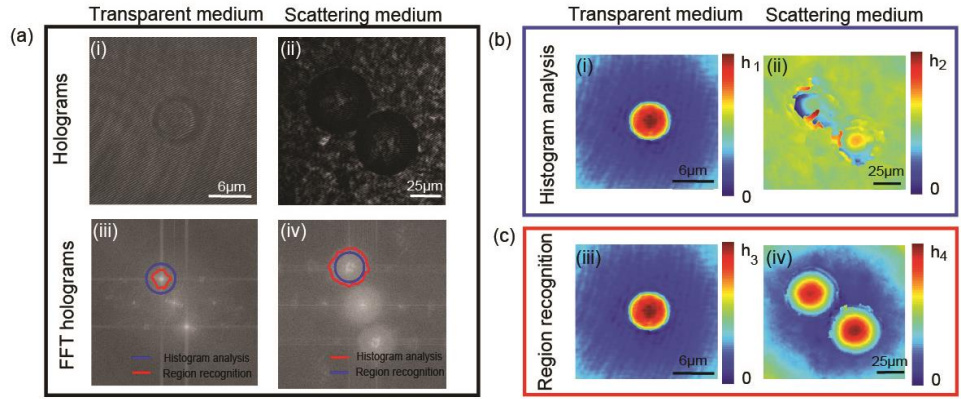

FIG. 3. Comparison of histogram analysis filtering technique [18] with region-recognition method under different sample conditions. (a), (i) and (ii) are the holograms of a 6 µm diameter microsphere through water and two 50 µm diameter microspheres through opaque peptide hydrogel. (iii) and (iv) are the FFT holograms of (i) and (ii) respectively. In (iii) and (iv), the red outlines represent the best filters from region recognition method and blue circles represent manual pre-cropping area used for histogram analysis. Reconstructed phase image with histogram analysis are shown in (b) and region recognition method (c) .Color bar $h_1$ = 6.7 µm Color bar $h_2$ = 50 µm Color bar $h_3$ = 6.7 µm, Color bar $h_4$ = 50 µm

The caveat of previous histogram analysis technique [18] is the requirement to first manually pre-cropping the area containing the first order frequency. The histogram of the area would then be analysed accordingly (shown in more detail in Appendix A1). The drawback of using such manual pre-cropping process is that it influences the quality of the final reconstruction from sample to sample. We illustrate this point by directly applying the same pre-cropping and histogram analysis to retrieve phase images of polymer microspheres in a transparent medium into a scattering medium. The reconstructed phase is then compared with the iterative threshold technique.

The purpose of this study is to prepare an automated DHM system that is suitable to study living cells develops into a living tissue. In thick samples, such as thick tissue culture, the effect of scattering is more pronounced and hence the reconstruction becomes more complex due to higher frequency noises. We therefore test the thresholding process on their performance sample with thick (~ 100 µm) optically scattering biological-relevant matrix (peptide-hydrogel). The scattering medium (Appendix Fig. A1) is prepared with 50 µm polymer microspheres (Duke Scientific) deposited onto a thin film of biological hydrogel matrix peptide [25] hydrogel (Fmoc-FRGDF). The peptide hydrogel has been used in thick cell growth culture for neuronal growth. The hologram of a polymer microsphere in water, Fig. 3.(a)(i), is significantly different from the hologram of hydrogel matrix [Fig. 3.(a)(ii)], as the latter displays a large background scattering with significant reduction in opacity. The scattering sample generates additional spatial frequency content that "blurs" the boundaries and "extends" the dimensions of three spatial frequencies components [FIG. 3.(a)(iv)]. That adds complexity to precise selection of first order frequency content for phase retrieval process. The results of the reconstructed phase maps from the histogram analysis technique, Fig 3. (b), and our approach, Fig. 3(c), show that the histogram analysis approach works in a standard transparent medium Fig. 3(b) (i). but clearly fails to extract the appropriate spatial frequency in the scattering medium as shown in Fig. 3 (b) (ii). The iterative methodworks for both samples as shown in Fig. 3(c) (i) and (ii). The comparison indicates that the filtering process is complicated by the broad distribution of spatial frequency. Previous histogram analysis techinique failed to obtain the desired reconstruction. While our approach, iterative thresholding and region recognition, successfully retrieved optimal shape of the filtering windows [blue curves in Fig. 3(a)(iii) and (iv)] and obtained accurate reconstruction of the microsphere through a diffusive medium [Fig. 3(c)]. A more detailed breakdown of the steps of both the histogram analysis technique and threshold increment determination are shown in Appendix A1. We also include another histogram analysis [17] in Appendix A2 that also fails to retrieve the spatial frequency of scattering medium.

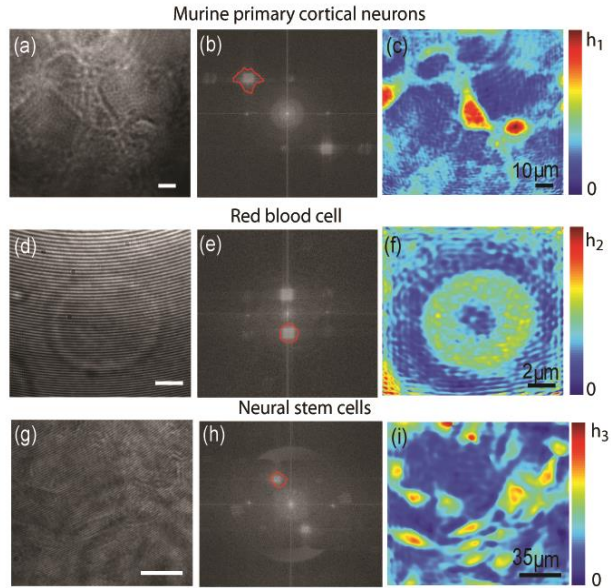

FIG. 4. Application of the region-recognition method under different experimental conditions. The first row depicts neuronal cortical cells (hologram (a), FFT image (b) and phase reconstruction (c)). The second row shows a single RBC as hologram (d), FFT image (e) and phase reconstruction (f). The last row presents C17.2 neural stem cell as hologram (g), FFT image (h) and phase reconstruction (i). The color bars in all phase reconstructions indicate $h_1=10$ µm, $h_2= 2.9$ µm, $h_3= 13.3$ µm .

After unwrapping and compensation, the remaining operation is to calculate the optical height of the sample based on wavelength $\lambda_0$ and the refractive index difference $\Delta n$ between the sample (polymer sphere: 1.59 or cell: 1.36) and the media (air: 1, water: 1.33 or oil: 1.51). The height differences within a sample can be plotted as a map using phase value φ(x,y).

$$h(x,y) = \frac{\varphi(x,y)\lambda_0}{2\pi\Delta n} \qquad (8)$$

To demonstrate the adaptability of our system to a range of experimental conditions we tested it using murine primary cortical neurons (cells that are adhesive, polymorphic, and relatively large), human RBCs (cells in suspension, biconcave and small) and microspheres (polystyrene, uniform structures of intermediate size). Figures 4(a), (b) and (c) show the hologram, FFT image and phase reconstruction of murine primary cortical neurons. The hologram was taken under 10X objective (N.A. 0.25) with fringes 45 degrees to the right in fine spacing (3 pixels per one fringe). Figures 4(d), (e) and (f) show the hologram, FFT image and phase reconstruction of a single RBC taken under high magnification (100 X, N.A 1.25 objective) with 15 pixels per one fringe. Figures 4(g), (h)and (i) show the hologram, FFT image and phase reconstruction of a single 6 μm polymer microsphere with fringes oriented inclined at 30 degrees with 8 pixels per one fringe. Figures 4(j), (k) and (s) show the hologram, FFT image and phase reconstruction of a C17.2 cell line taken under 10X objective (N.A. 0.25) with fringes 30 degrees to the right in fine spacing with 4 pixels per one fringe. These results show that the iterative thresholding and region-recognition method can process a diverse range of shapes of spatial frequencies.

The entire numerical processing is implemented in a MATLAB program with a graphical user interface (GUI) as shown in Fig. 5(a). In the GUI, the hologram will be displayed at the right side of control panel with variable settings on gain and exposure time (based on camera settings). A region of interest is selected, cropped and displayed. The phase reconstruction is shown with two views: perspectives and plan.

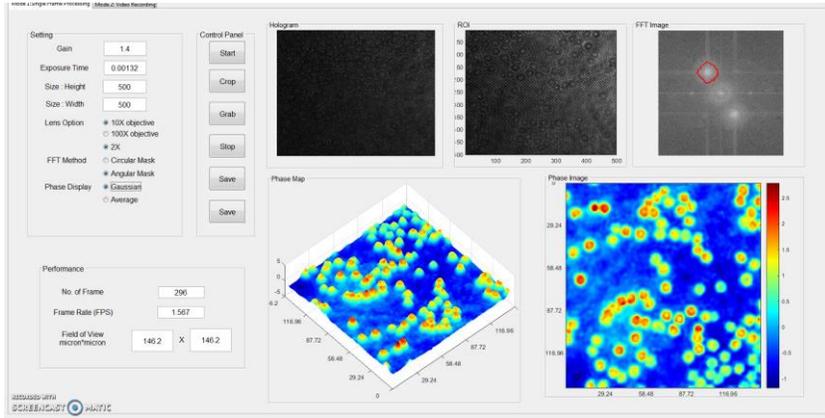

FIG. 5. Layout of graphical user interface: red blood cells recorded with 10X NA 0.25 microscope objective (Visualization 1)

## 4. Region-recognition for classification of infected RBCs

Analysis of cellular characteristics are often restricted to one of two scenarios: either averages of cell populations are examined (missing any heterogeneity that might be present within the population) or individual cells are studied (with the disadvantage of limited numbers and restricted in-vivo compatibility). Furthermore, most analytical imaging methods involve the addition of labels (e.g. fluorescent tags, antibodies, chemical dyes), which potentially perturb the examination. Our system facilitates the classification of cellular populations without the aid of labels rapidly by retrieving phase values for individual cells.

This is critical for various biological applications. For example, RBCs are the most abundant cell type in the human body [26]. Mutations in RBCs are very common and some have serious implications for the wellbeing of the carrier (e.g. sickle cell anemia, beta thalassemia, ovaloycytosis to name only a few [27]). Moreover, infections (e.g. malaria, babesiosis [28]) and drugs (e.g. iodinated contrast media [29]) can also influence the characteristics of RBCs and the analysis of the characteristics is important for diagnostics, disease outcome and research.

Here the region-recognition program is expanded toward characterization of RBCs. Fig.6 illustrates the key steps (thresholding, perimeter location) used to extract individual RBCs. The first step is to read image and apply a Gaussian filter for thresholding (Otsu's method [30]). Next, holes are filled to remove small objects and extract perimeter pixels. The third step is to get the center regions of individual cells and do the same processing of second step with these center regions. The fourth step is to inverse image intensity. The fifth step is to separate objects using watershed transform and measure the pixel areas of each object in the combined binary image. The final step is to label cell regions and remove false cell detections based on the pixel areas of objects. The phase value of each identified cell is saved into arrays for further analysis.

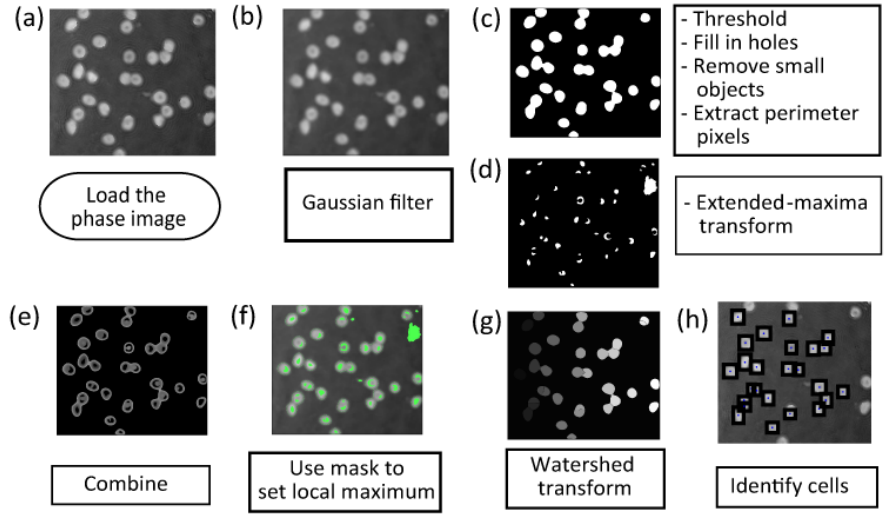

FIG. 6. Region-recognition to extract individual RBCs. (a) load and (b) apply a Gaussian filter for thresholding. (c) voids are filled, small objects removed and perimeter pixels extracted. (d) maxima transform performed and apply step (e) combined the (c) and (d), (f) Invert image intensity and set local maximum in mask. (g) separate objects using watershed transform and (h) measure the pixel areas of each object in the combined binary image. (h) The final step is to label cell regions and remove false cell detections based on the pixel areas of objects.

The distribution of phase values represents the height variation of individual cells that has statistical significant difference between malaria parasite infected and uninfected RBCs. Previous studies shown that infected RBCs possess a distinctly different morphological deformation when compared to the uninfected RBCs [31]. Figures 7(a) and (b) are the SEM pictures of malaria parasite infected and uninfected RBCs respectively, while Figs. 7(c) and (d) are the phase reconstructions of an infected and an uninfected RBCs, which were taken and reconstructed from our system with high magnification objective lenses (100X, N.A 1.25). The color bar beside shows the phase values by unit of radians. Seen from the figures, the phase maps present the consistency with the SEM pictures, where infected RBC has anomalous shape while uninfected RBC shows normal biconcave shape.

In order to pick out the morphological differences between infected and uninfected RBCs, the phase value over the cell surfaces is selected out to plot the histograms. The black dots in Figs. 8(a), (b) are the histograms of normalized phase values over RBCs shown as thumbnails respectively. The histogram indicates how height of a cell distributes over the 2D plane. The solid curves in Figs. 8(a), (b) are Gaussian fittings of the histograms. Quantification of the phase is inferred from the full width at half-maximum (FWHM) of the fitted curve of the phase distribution provides an indication of the spatial distribution of the phase (height of a cell) variation over an entire cell as shown in Fig. 8 (a), (b) shaded region.

Figure 8(c) shows the distribution of FWHMs of the spatial distribution of the phases on 12 infected and 12 uninfected RBCs. The trends indicate a marked difference between infected and uninfected RBCs. The infected RBCs is most likely have lower FWHMs (the histogram is more centralized), which means the differences in cell height are not major. Therefore, the blue line in Fig. 8(c) is the histogram of FWHMs of infected RBCs. It shows most of infected RBCs have FWHMs at below 0.1, while uninfected cells (red line) tend to have higher FWHMs (0.25 to 0.3). This shift indicates that the height of uninfected RBCs changes largely spatially. In other words, the uninfected RBCs have broader height (phase) distribution. It can be explained by the fact that the uninfected RBCs have thinner central height. Therefore, the height difference between the centre and the periphery is larger than that of infected RBCs where the volume of the centre part is occupied by the parasite.

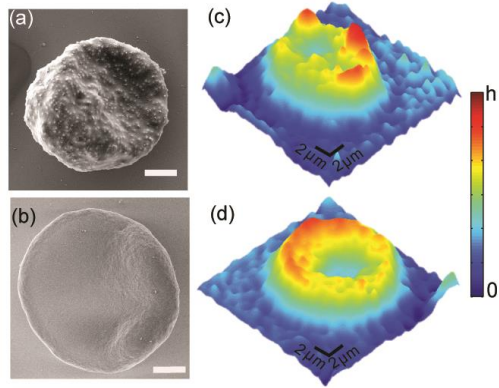

FIG. 7. Example images of red blood cells taken with scanning electron microscope (a), (b) and DHM (c), (d). Infected (a),(c) and non-infected (b) and (d). h =2.77 µm

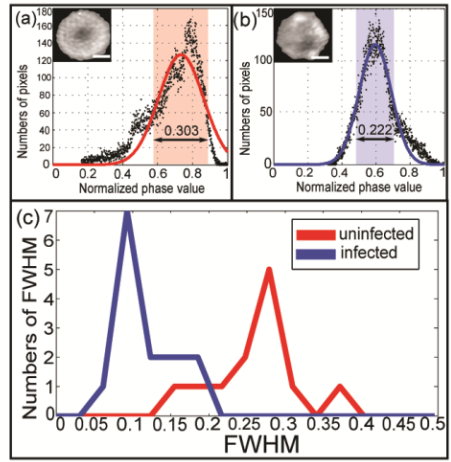

Fig. 8. Phase distribution analysis of malaria infected RBCs versus uninfected RBCs. (a)Histogram of phase value over an infected RBC surface shown in inset 2µm scale bar. (b)Histogram of phase value over an uninfected RBC surface, which is shown as thumbnail with 2µm scale bar. (c) Histograms of FWHMs of 12 uninfected RBCs' phase distribution (red) and the histogram of FWHMs of 12 infected RBCs' phase distribution (blue). This shift between two histograms indicates that uninfected RBCs have broader height (phase) distribution.

## 5. Conclusion

DHM relies heavily on accurate execution of numerical processes to reconstruct the phase of the biological sample. While off-axis DHM technique is perhaps the fastest holographic imaging technique, the accurate filtering of the spatial frequency can affect the final reconstructed image. We demonstrated that the combination of region-recognition and iterative thresholding conducted in a single loop can aid in selecting the optimal spatial frequency components with marginal delay (~0.1 to 0.3s). The effectiveness of our process is illustrated in the comparison study where the iterative approach clearly identifies reconstructed microspheres embedded in opaque hydrogel matrix (~ 100 µm) has addressed the use of the technique in thick samples. More importantly, as compared with past techniques, we showed that this technique can operate well in both transparent and turbid system, thus providing potential flexibility for imaging live cell activites [32]. In the next step, we will aim to conduct live cell imaging in confluent cell cultures that will aid in the study of cellular process in engineered tissue.

**Acknowledgements**

XF He is grateful for the sponsorship from China Scholarship Council. W. M. Lee acknowledges support from the FERL start-up funds from Research School of Engineering, Australian Research Council Early Career Researcher Award (DE160100843) and ANU Major Equipment grant (15MEC36). D. R. Nisbet is supported by an NHMRC Career Development Fellowship (APP1050684). C. V. Nguyen is supported by the Australian Research Council Centre of Excellence for Robotic Vision (project number CE140100016), http://www.roboticvision.org. The authors acknowledge the expertise and facilities of the Australian Microscopy and Microanalysis Research Facility at the Centre for Advanced Microscopy (CAM), ANU.

**Appendix A1 – Histogram analysis method 1 [18]**

Here we repeated the histogram analysis[18] procedures as shown in Fig.3 To test the technique's ability to adapt, we prepared the sample under a highly diffused medium, a thin film of opaque peptide hydrogel (Fmoc-FRGDF), which shown in Fig. A1(a). Figure A1(b) shows the detailed breakdown of the histogram selection approaches against the automated region recognition approach as described in Fig. A1 (c). The histogram analysis process requires a Gaussian curve fit followed by 2 sets of differentiation to retrieve the inflexion point. While it works well in standard samples, it does not operate well on scattering sample as shown in Figure A1(b) it shows that the technique cannot accurately retrieve the appropriate phase. For our region recognition, the step increment of the threshold level is set as 1% of GTL and gradually decrease the increment threshold level. During the process, the filtering window progressive enlarges and capture the necessary. When the increment is reduced down to 1% of GTL (GTL=0.4627), the reconstruction outcome was shown to display the phase of the microsphere. The caveat of the iterative process is that the first order and zero order must not overlap, which is the limitation of all off-axis DHM systems.

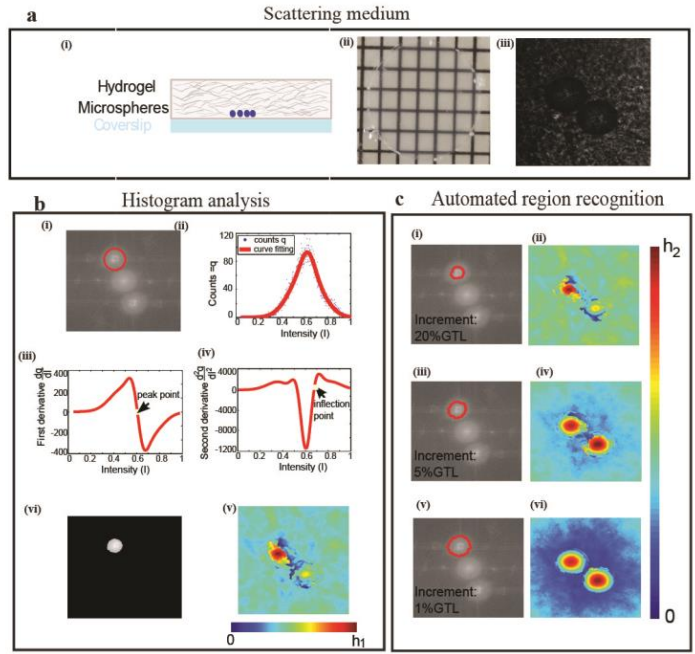

Fig A1 shows the detailed comparison of the phase retrieval process of microsphere embedded n a scattering medium (a) – (i) schematic, (ii) image of scattering medium taken from plan view with grid (iii) acquired hologram. b (i) shows the circle binary filter applied to FFT holograms of the sample. (ii) the histogram of intensity distribution after manually filtering the spatial frequency and Red curve show a Gaussian fit (sum of two Gaussians functions) of the distribution. (iii) and (vi) are 1st and 2nd derivative of the fitting curve in (ii). Inflexion point (intensity =0.67) in iv is used to retrieve the threshold level to obtain the filter in (v) and (vi) and the reconstruction, which insufficiently retrieves the microsphere. (c) increment determination process in automated region recognition method. Figures c(i), (iii), (v) show the different filter windows and (ii),(iv),(vi) are the corresponding reconstructed phase of the microsphere. The threshold windows are created by setting different threshold increment in step (2), Fig.2 (j). Each threshold increments goes from 20% - (i), (ii), 5% - (iii), (iv) and 1% - (v), (vi) of GTL. The results show that 1% of GTL achieves an optimal filter window to retrieve an accurate reconstruction of the microsphere through a diffusive medium. Color bar $h_1 = h_2 = 50$ µm

**Appendix A2– Histogram analysis method 2[17]**

In this section we implemented a second reported histogram techniques [17]. The crux of the technique is to select a "second" peak in the histogram count. As shown in Figure.A2, their technique operates by applying a negative Laplacian [20] to suppress zero order of spectrum (FFT hologram in our case) and then plotting the histogram of the processed spectrum to locate the "peaks" as shown Fig.A2 (b) and (f). The difference of the intensity of two peaks (intensity of the background and maximum gray-scale point) is then used to allocated the threshold. However, when applying this method to the same samples in Fig. 3, there was no sufficient information to clearly locate the second peak (maximum gray-scale point). Figure A2 (b) and (f) are the histogram of Fig.A2 (a) and (e). As shown in (b), a second peak in intensity at 0.72 that corresponds to a difference in the threshold to be 0.01 (0.72-0.61=0.01). **Such low threshold essential covers the entire FFT hologram in Fig.A2(a) and hence the filter window covers all the spectrum [red rectangle in Fig A2(a)].** This is the same for Fig A2(e). Figs. A2 (c), (d) and (g), (h) are the phase and amplitude reconstructions of two samples under this technique. This method failed to work in locating the right threshold.

As the result, we needed to manually intervene and insert threshold values by visual judgement of the retrieved phase. The thresholds now becomes 0.79 and 0.67 as shown in Fig A2(i) and A2(k) respectively with desired outcome in Fig.A2 (j) and (i). By relying on the previously reported histogram technique [18], this would mean that the second peak needs to lie at a position larger than the normalize value of 1.

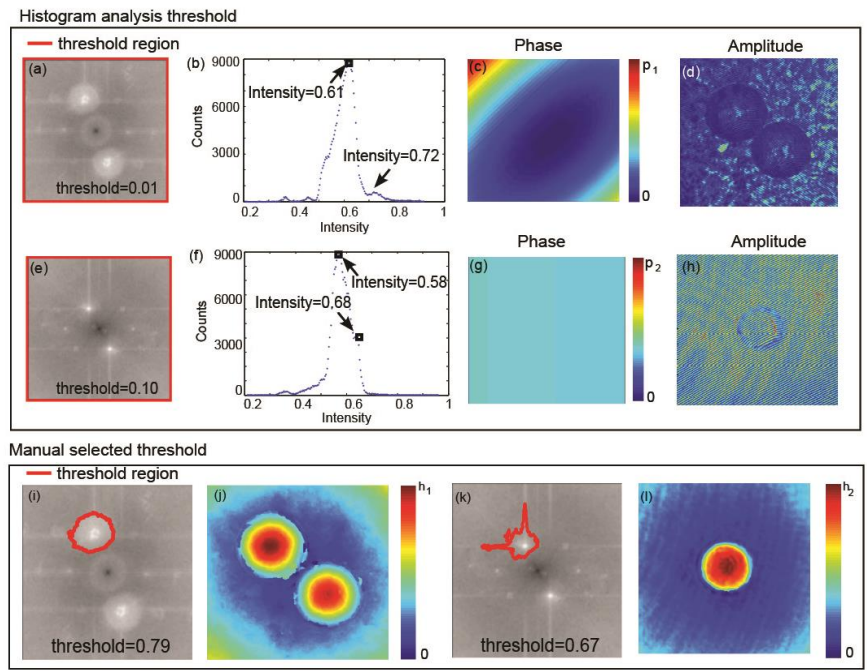

Fig A2 shows the detailed comparison of the phase retrieval process of the same sample in Fig 3. The detailed histogram analysis in the previous technique [18] are shown in the top: (a) and (e) are the processed spectrums, in which the red rectangles are the filter windows based on histogram analysis. (b) and (f) are the histograms of (a) and (e) respectively. (c) and (d) are the phase and amplitude reconstructions of (a). (g) and (h) are the phase and amplitude reconstructions of (e). The manual determination of threshold is shown downside: (i) shows the filter window when threshold is 0.79 and (j) is the phase reconstruction of it. (k) shows the filter window when threshold is 0.67 and (l) is the phase reconstruction of it. Color bar $h_1$ = 50 µm. Color bar $h_2$ = 6.7 µm. Color bar $p_1$=1.0e-4 rad. Colorbar $p_2$= 0.01 rad.